# A Novel Hybrid Machine Learning Model for Rapid Assessment of Wave and Storm Surge Responses Over an Extended Coastal Region

## Saeed Saviz Naeini[a], Reda Snaiki[a,*]

[a] Department of Construction Engineering, École de Technologie Supérieure, Université du Québec, Montréal, QC, H3C 1K3, Canada

*Corresponding author. Email: reda.snaiki@etsmtl.ca

**Abstract:** Storm surge and waves are responsible for a substantial portion of tropical and extratropical cyclones-related damages. While high-fidelity numerical models have significantly advanced the simulation accuracy of storm surge and waves, they are not practical to be employed for probabilistic analysis, risk assessment or rapid prediction due to their high computational demands. In this study, a novel hybrid model combining dimensionality reduction and data-driven techniques is developed for rapid assessment of waves and storm surge responses over an extended coastal region. Specifically, the hybrid model simultaneously identifies a low-dimensional representation of the high-dimensional spatial system based on a deep autoencoder (DAE) while mapping the storm parameters to the obtained low-dimensional latent space using a deep neural network (DNN). To train the hybrid model, a combined weighted loss function is designed to encourage a balance between DAE and DNN training and achieve the best accuracy. The performance of the hybrid model is evaluated through a case study using the synthetic data from the North Atlantic Comprehensive Coastal Study (NACCS) covering critical regions within New York and New Jersey. In addition, the proposed approach is compared with two decoupled models where the regression model is based on DNN and the reduction techniques are either principal component analysis (PCA) or DAE which are trained separately from the DNN model. High accuracy and computational efficiency are observed for the hybrid model which could be readily implemented as part of early warning systems or probabilistic risk assessment of waves and storm surge.

**Keywords**: Storm Surge, Significant Wave Height, Deep Learning, Deep Autoencoder

## 1. Introduction

Hurricanes are among the most devastating natural hazards which can cause significant damage and loss of life and property (Chen et al., 2008; Lin et al., 2010). Coastal communities are specifically extremely vulnerable to hurricane-induced surge and waves (Wamsley et al., 2009). This vulnerability has been recently illustrated with several major and costly hurricanes (e.g., Harvey 2017, Dorian 2019, Ian 2022). With the increasing coastal population, sea-level rise, and climate change effects, both economic losses and human fatalities are expected to increase disproportionately over hurricane-prone regions (Zhang et al., 2000). For instance, an increase of the hurricane-induced annual losses in the United States is expected to reach $39 billion in the coming years compared to current values of $28 billion (Dinan, 2016). Hence, accurate and efficient modeling of storm surge and waves is critical to support effective risk-responsive decision-making and mitigation of hurricane-induced losses.

Storm surge is driven by surface wind stress and pressure gradient. The magnitude of the storm surge is also affected by several parameters, including the storm size, basin geometry and bathymetry (Irish and Resio, 2010; Lin and Chavas, 2012). On the other hand, the waves are mainly driven by the momentum transfer through wind stress and are affected by several factors such as the ocean currents (Fan et al., 2009). Several modeling approaches have been proposed to simulate both storm surge and waves. This includes the simplified empirical and statistical approaches (e.g., Rao and Uztilldar, 1966; Bretschneider, 1967). However, the use of preselected basis functions may not always guarantee accurate simulations especially for highly nonlinear systems (Sztobryn, 2003; Thomas and Dwarakish, 2015). In addition, these models are usually developed for restricted locations and are not applicable for other regions.

To improve the accuracy of the simplified empirical/statistical models, considerable efforts have been made, over the past few decades, to solve numerically the nonlinear governing equations of both storm surge and wave. For instance, the Advanced CIRCulation (ADCIRC) is a high-fidelity high-resolution numerical model based on finite-element which solves the shallow-water equations and provides the water levels along with the ocean currents values (Luettich and Westerink, 2004). The Simulating WAves Nearshore (SWAN) is an example of a high-fidelity phase-averaged numerical model for generating the ocean waves which solves the action balance equation (Booij et al., 1999). For practical implementation, the ADCIRC and SWAN models are simultaneously solved to account for the inherent coupling between storm surge and waves. Specifically, the SWAN model generates the radiation stresses which are passed to ADCIRC. The latter generates the water levels along with the currents which are required by the SWAN model. Other numerical models have also been developed such as the Sea Lake and Overland Surges from Hurricanes (SLOSH) (Jelesnianski, 1992) for storm surge prediction and the Steady-State Spectral Wave Model (STWAVE) (Smith et al., 1999) for wave modeling. However, although the high performance of the numerical models has been recognized widely, their high computational cost makes it difficult, and even impractical to use them for probabilistic analysis, risk assessment or real time/rapid predictions.

To overcome the computational burden of high-fidelity numerical models, data-driven techniques offer another alternative for rapid assessment of hurricane-induced storm surge and waves. Data-driven techniques identify the inherent relationship between the predictors (e.g., storm parameters) and the predicted (e.g., storm surge and wave) based on the available historical or numerical (or a combination of both) data. In particular, machine learning (ML) techniques have drawn attention in recent years due to their efficiency and robustness (Wu and Snaiki, 2022). For example, Bezuglov et al. (2016) developed a feed-forward artificial neural network (ANN) for storm surge prediction in North Carolina. Lee (2006) trained an ANN algorithm to predict storm surge in Taiwan with a one-hour lead time based on meteorological data. Kim et al. (2015) trained an ANN based on synthetic hurricane data to predict the time-dependent storm surge at few locations in southern Louisiana. Hashemi et al. (2016) proposed two ML models, namely ANN and support vector machine (SVM), for the prediction of peak storm surge for Rhode Island. The ANN model outperformed the SVM algorithm. Several other studies have also employed the ANN algorithm for the prediction of peak or time-dependent storm surge (e.g., Bajo and Umgiesser, 2010; Chen et al., 2012; French et al., 2017; Al Kajbaf and Bensi, 2020; Ramos-Valle et al., 2021; Lockwood et al., 2022). Moreover, Kriging metamodels (also known as Gaussian process), have been successfully used in various studies to model storm-induced surge responses (e.g., Jia and Taflanidis, 2013; Zhang et al., 2018; Kijewski-Correa et al., 2020; Nadal-Caraballo et al., 2020; Plumlee et al., 2021; Kyprioti et al., 2021, 2022). For example, Zhang et al. (2018) proposed a kriging model for storm surge prediction using the North Atlantic Comprehensive Coastal Study (NACCS) database. In addition, they have presented an adaptive sequential design of experiment for selecting storms to enhance the model's predictive capacity. Similarly, Kyprioti et al. (2021) selected a Gaussian Process regression model to estimate the peak storm surge response around Delaware Bay based on synthetic storms retrieved from the FEMA Region 3 Coastal Storm Surge Study (Hanson et al., 2013). Other surrogate models (e.g., moving least squares and recurrent neural network) have also been proposed in other studies for the prediction of the peak or time-dependent storm surge (e.g., Irish et al., 2009; Taflanidis et al., 2013; Tadesse et al., 2020; Igarashi & Tajima, 2021; Adeli et al., 2022; Bai & Xu, 2022). Similarly, hurricane waves have also been predicted (e.g., significant wave height) using ML techniques (e.g., Berbić et al., 2017; Callens et al., 2020; Fan et al., 2020; Meng et al., 2021; Song et al., 2022; Gao et al., 2023).

Although significant contributions have been made to develop advanced ML models, it is still extremely challenging to predict storm surge and waves over an extended region using a single ML model. Training several ML models corresponding to each location (or a group of locations) is an alternative approach, however, it is not practically feasible due to the high computational demand and storage cost required to handle such high-dimensional space. Therefore, dimensionality reduction techniques have been integrated to identify low-dimensional feature space from the original high-dimensional space (Van Der Maaten et al., 2009). For example, Jia and Taflanidis (2013) utilized first the principal component analysis (PCA) technique to deal with the high-dimensionality of the output vector (e.g., peak storm surge), then employed kriging metamodeling for the approximation of hurricane surge/wave responses. Similarly, Jia et al. (2016) proposed a kriging-based model to predict the peak and time-dependent storm surge over a large area where PCA has been used as a dimensionality reduction technique. Lee et al. (2021) combined a convolutional neural network (CNN) with PCA and k-means clustering to rapidly predict peak storm surge from landfalling and bypassing tropical cyclones. Other studies have also implemented the PCA technique as a dimensionality reduction tool combined with data-driven techniques (e.g., Bass and Bedient, 2018; Kyprioti et al., 2021, 2022). Recently, Saviz and Snaiki (2022) used deep autoencoder (DAE) instead of PCA to effectively capture the nonlinearities within the data and combined it with several ML models including ANN, random forest regressor and gradient boosting regressor. The simulation results indicated that the DAE-based model outperformed the PCA-based model since it is more suitable for nonlinear systems. Although, the use of a dimensionality reduction technique (e.g., PCA or DAE) alleviates the computational burden, it comes with no guarantee that the identified latent space is the optimal one for the data-driven technique. Therefore, the latent space might not necessarily preserve the intrinsic dynamics which the data-driven technique will seek to identify from the input to the output. Hence, training the data-driven and dimensionality reduction techniques separately has several limitations (Champion et al., 2019).

In this study, a novel hybrid model is developed for the rapid prediction of hurricane-induced storm surge and waves over an extended coastal region. Specifically, the hybrid model couples both a deep autoencoder (DAE) and a deep neural network (DNN) which are trained simultaneously. While DAE identifies a low-dimensional representation of the high-dimensional spatial system, the DNN maps the storm parameters (e.g., central pressure deficit and radius of maximum wind) to the obtained low-dimensional latent space. A unique weighted loss function is designed to train the hybrid model to encourage a balance between DAE and DNN training and achieve the best accuracy. A case study related to the simulation of peak storm surge/significant wave height over critical regions within New York and New Jersey is selected to demonstrate the high performance of the proposed model. The training/testing data is retrieved from the North

Atlantic Comprehensive Coastal Study (NACCS). Additionally, the proposed framework is compared with two decoupled models consisting of a dimensionality reduction technique (PCA and DAE) and a regression model based on DNN which are trained separately.

## 2. Methods

### 2.1. Deep neural networks

High-fidelity numerical models have significantly advanced the analysis of hurricane-induced hazards (e.g., wind, rain, and storm surge). However, these models are computationally expensive; therefore, it is challenging to apply them for either near real-time forecasts or risk analysis which requires many simulations. To overcome the inherent limitations of high-fidelity numerical models, machine learning (ML) algorithms could be used for such applications (Jia et al., 2016; Snaiki and Wu, 2020). ML algorithms aim to discover a mapping function between the inputs (e.g., hurricane parameters) and outputs (e.g., storm surge) by minimizing a cost function. The necessary data required for training and testing the ML model can be retrieved from various sources such as high-fidelity numerical simulations, experimental setups and field measurements. Among many ML models, artificial neural networks (ANN) are widely utilized due to their superior capabilities in capturing the inherent nonlinearities in the data (Grossberg, 1988). Further discussion on the ANN models is provided in Appendix A. To improve the model predictions, and effectively represent the hidden nonlinear patterns within the data, deep neural networks (DNN) have been proposed as illustrated in Fig. 1. Although DNNs are quite similar to ANN models, they are characterized by a deep architecture consisting of several hidden layers (Sze et al., 2017). In this study, DNN models have been utilized to predict the hurricane-induced storm surge and significant wave height responses for both landfalling and bypassing storm scenarios.

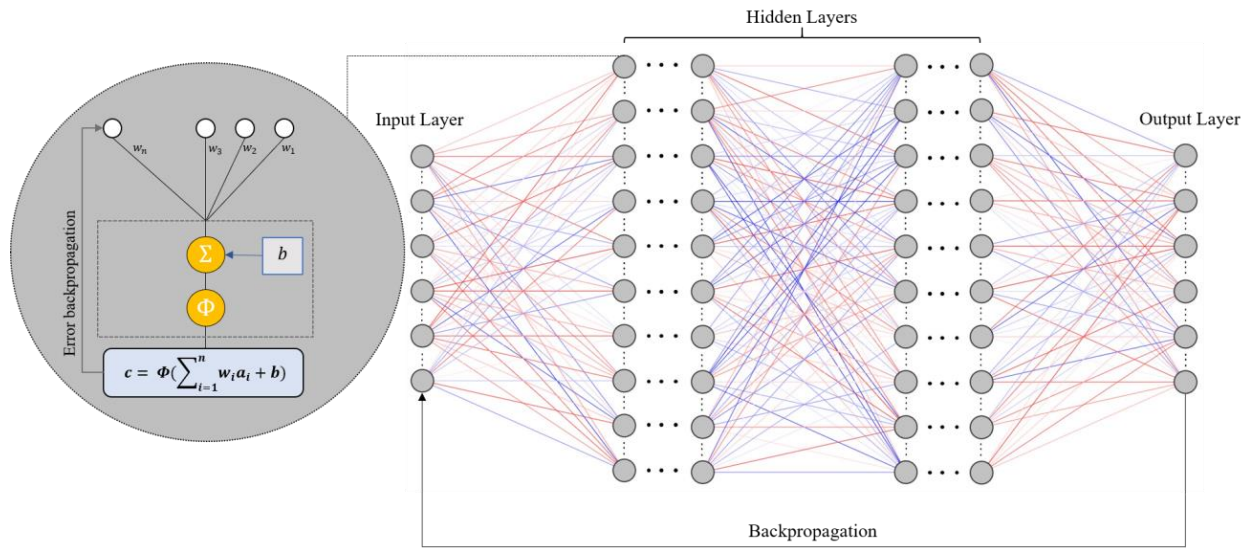

**Fig. 1.** Typical architecture of deep neural networks (DNNs) [further details are presented in Appendix A]

In this study, six hurricane parameters are taken as the DNN inputs, namely the reference latitude ($LAT$) & longitude ($LON$) as defined by the NACCS study for both landfalling and bypassing storms, heading angle ($\theta$), central pressure deficit ($C_p$), translation speed ($V_f$), and radius of maximum winds ($R_{max}$). These parameters form the input $\boldsymbol{x}$ of the DNN model which can be written as the following:

$$\boldsymbol{x} = [C_p(\text{hPa}), \quad \theta(°), \quad R_{max}(\text{km}), \quad LAT(°), \quad LON(°), \quad V_f(km/h)]^T \tag{1}$$

It should be noted that the heading direction ($\theta$) is measured clockwise from North, where 0° indicates a storm track heading North and minus sign representing storm directions towards the west. Furthermore, the output vector $\boldsymbol{y}$ represents the peak storm surge/significant wave height values over the selected locations (or a cluster of locations).

However, even with the most advanced ML algorithms, it is extremely challenging to predict hurricane-induced hazards over a vast coastal region using a single ML model due to the high dimension of the output vector (e.g., storm surge) (Jia and Taflanidis, 2013). Training several ML models corresponding to each location (or a group of locations) is an alternative approach to improve the prediction accuracy. However, it is not practically feasible due to the high computational demand and storage cost required to handle the high-dimensional output vector.

## 2.2. Dimensionality reduction

Hurricane response (e.g., storm surge) is usually estimated over a large coastal region leading to a high-dimensional output vector. Therefore, it is impractical to train a data-driven model over this high-dimensional space. To deal with high-dimensional data, several dimensionality reduction techniques have been developed. Specifically, they identify a low-dimensional feature space from a high-dimensional space (Van Der Maaten et al., 2009). Therefore, the number of attributes is usually significantly reduced, especially when a strong correlation between the original features exists. Once a low-dimensional space is identified, reduced order models can be implemented over the low-dimensional representation.

Dimensionality reduction algorithms can be generally classified into linear and nonlinear techniques. Linear dimensionality reduction techniques map high-dimensional data to lower-dimensional space using linear functions (Portnova-Fahreeva et al., 2020). Principal component analysis (PCA) is one of the most popular linear dimensionality reduction techniques. It aims to identify orthogonal directions (also denoted as principal components) where the variance of the projected data is maximized (Abdi and Williams, 2010). Although linear reduction techniques are easy to implement, they do not properly handle complex nonlinear structures. Therefore, nonlinear dimensionality algorithms have been proposed in order to obtain representative low-dimensional spaces from the original high-dimensional data (Van Der Maaten et al., 2009). Among the nonlinear dimensionality reduction techniques, autoencoders have recently gained popularity given their ability to significantly compress highly nonlinear data. Autoencoders are unsupervised artificial neural networks which consist of an encoder, a code (i.e., latent space) and a decoder (Liou et al., 2014) as illustrated in Fig. 2. The encoder $'\varphi(.)'$ task is to transform the high-dimensional input space $'\boldsymbol{y}'$ into a lower-dimensional space $'\boldsymbol{z}'$, also denoted as the latent space. This leads to an $l$-dimensional latent output $\boldsymbol{z} \in \mathbb{R}^l$ out of $n$-dimensional inputs $\boldsymbol{y} \in \mathbb{R}^n$ such that $\boldsymbol{z} = \varphi(\boldsymbol{y})$ where $l \ll n$. On the other hand, the decoder $'\psi(.)'$ converts the identified latent space into the output space. The latter is given as $\hat{\boldsymbol{y}} = \psi(\boldsymbol{z})$ and should correspond to the reconstructed input space if the model is well-trained.

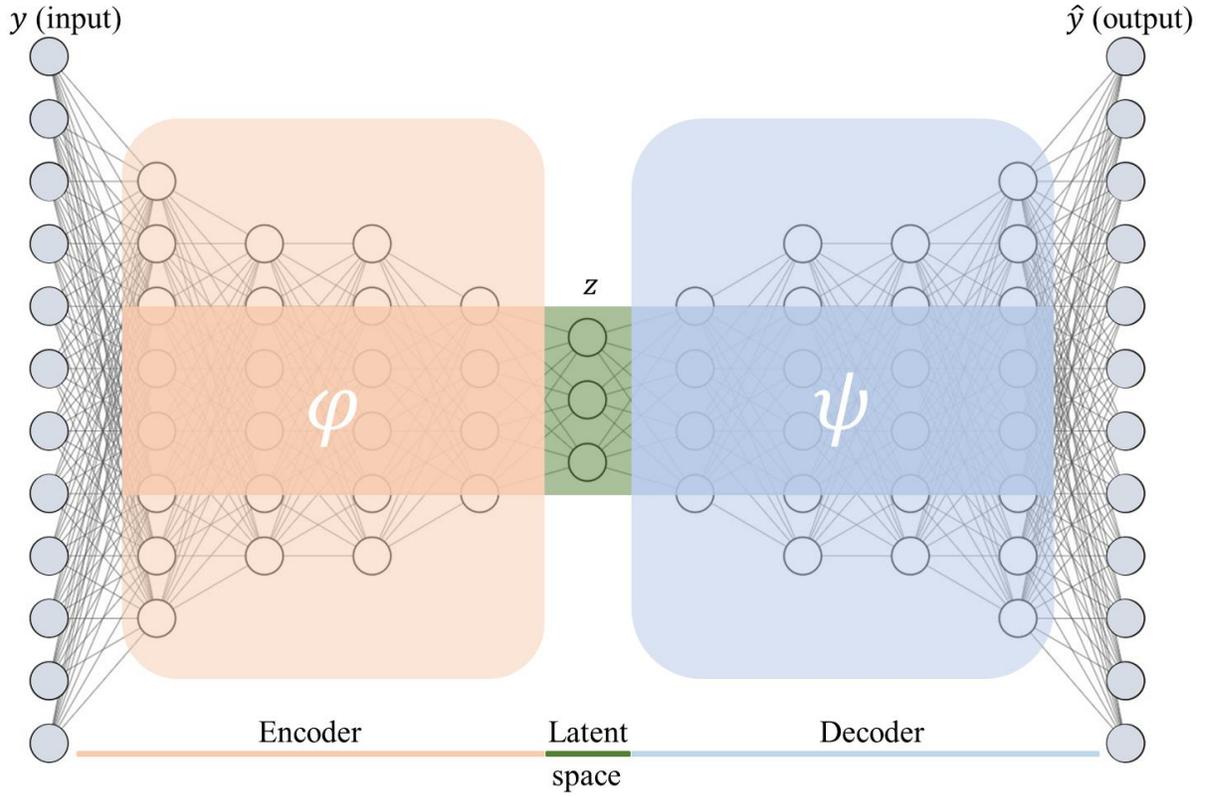

**Fig. 2.** Architecture of typical autoencoders

Autoencoders are capable of identifying nonlinear patterns within the data based on the employed nonlinear activation functions. Hence, PCA can be considered as a special case of autoencoders in which linear activation functions are used (Wetzel, 2017). When autoencoders have several hidden layers, they are commonly referred to as deep autoencoders. In this study, the $'\boldsymbol{y}'$ vector which represents the peak storm surge (or significant wave height) response over an extended area is typically a high-dimensional vector, and $'\boldsymbol{z}'$ corresponds to its low-dimensional representation over the latent space.

As indicated in Fig. 3, once a low-dimensional space $'\boldsymbol{z}'$ has been identified from a high-dimensional output space $'\boldsymbol{y}'$ (i.e., peak storm surge or significant wave height over a large area), a reduced order model (or several models) can be trained to map the input parameters $'\boldsymbol{x}'$ (i.e., hurricane parameters) to the low-dimensional output vector $'\boldsymbol{z}'$. Then, a reconstruction step is needed (e.g., decoder) to transform the obtained vector to the original space.

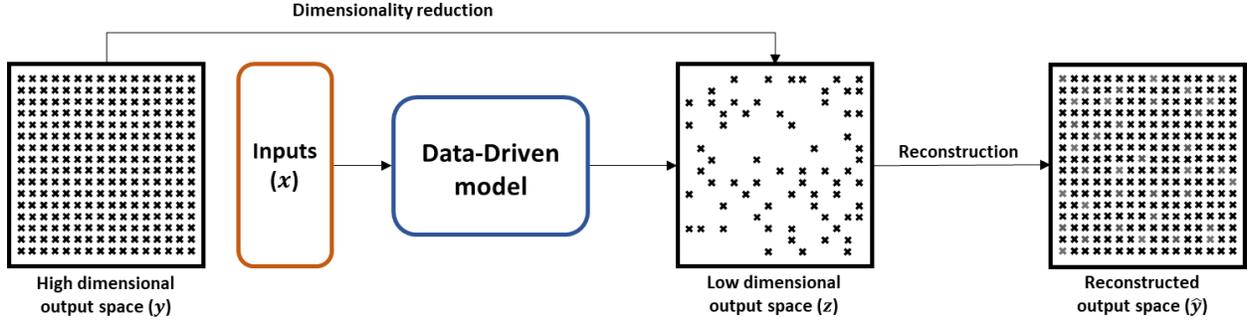

**Fig. 3.** Schematic implementation of a data-driven model and a dimensionality reduction technique

Although the previous methodology alleviates the computational burden, it comes with no guarantee that the identified latent space is the optimal one for the regression model which will map an input vector to the identified latent output vector. Therefore, the latent vector might not always preserve the intrinsic dynamics and nonlinearities which the regression model will seek to identify from the input to the output. Hence, training the autoencoder model and the regression model independently has several limitations (e.g., Champion et al., 2019).

## 2.3. Proposed model

A novel hybrid model is developed in this study to simultaneously identify a low-dimensional representation of high-dimensional systems while mapping input parameters to the obtained low-dimensional latent space. This model determines the required low-dimensional latent space for the development of a suitable regression model to achieve the best accuracy. Standard approaches which decouple the dimensionality reduction and regression models are limited, and might even fail, because they do not necessarily provide the right low-dimensional space which will still ensure that the intrinsic nonlinear relationship between the input and output (i.e., latent space) is preserved. Therefore, by simultaneously optimizing the dimensionality reduction algorithm and the regression model, a robust, generalizable, and meaningful model can be obtained by promoting a balance between the capabilities of dimensionality reduction and regression techniques. The proposed model consists of a deep autoencoder (DAE) and a deep neural network (DNN), dubbed here as DAE-DNN. While the DAE model enables the discovery of low-dimensional representation from a high-dimensional space (i.e., peak storm surge or significant wave height over a large area), the DNN model identifies the nonlinear relationship between the input parameters (i.e., the six storm parameters of Eq.1) and the identified latent space (i.e., latent output). Once the model has been trained, the DNN model will predict the desired values (i.e., peak storm surge or significant wave height) over the latent space, then the decoder will reconstruct the full system to provide a prediction over the entire area. To train the hybrid model, a weighted loss function is designed to encourage a balance between DAE and DNN training to achieve the best accuracy while accounting for the system constraints.

To illustrate the proposed hybrid model derivation, consider the following system which predicts a vector $\boldsymbol{y}$ (i.e., peak storm surge or significant wave height over a large area) given an input $\boldsymbol{x}$ (i.e., the six storm parameters of Eq.1) through a nonlinear function $f$ (unknown):

$$f(\boldsymbol{x}) = \boldsymbol{y} \tag{2}$$

Since the goal is to identify a function $\hat{f}$ which approximates the true function $f$, several data-driven techniques could be used (e.g., DNN). However, the high-dimensional output vector $\boldsymbol{y} \in$ $\mathbb{R}^n$ makes it extremely challenging to identify a suitable model (where $n$ denotes the number of geographical locations in this study). Therefore, a low-dimensional vector $\boldsymbol{z} \in \mathbb{R}^l$ should be first determined. This latent output can be identified using the encoder part of the DAE model ($\boldsymbol{z} = \varphi(\boldsymbol{y})$). A DNN model ($g$) can be then trained to map the input $\boldsymbol{x}$ to the latent output providing an approximation $\hat{\boldsymbol{z}}$ to the real $\boldsymbol{z}$ (i.e., $g(\boldsymbol{x}) = \hat{\boldsymbol{z}}$). In the last step, the decoder will reconstruct the original space $\hat{\boldsymbol{y}} = \psi(\hat{\boldsymbol{z}})$. The architecture of the proposed hybrid model is illustrated in Fig. 4 where both DAE and DNN are trained simultaneously.

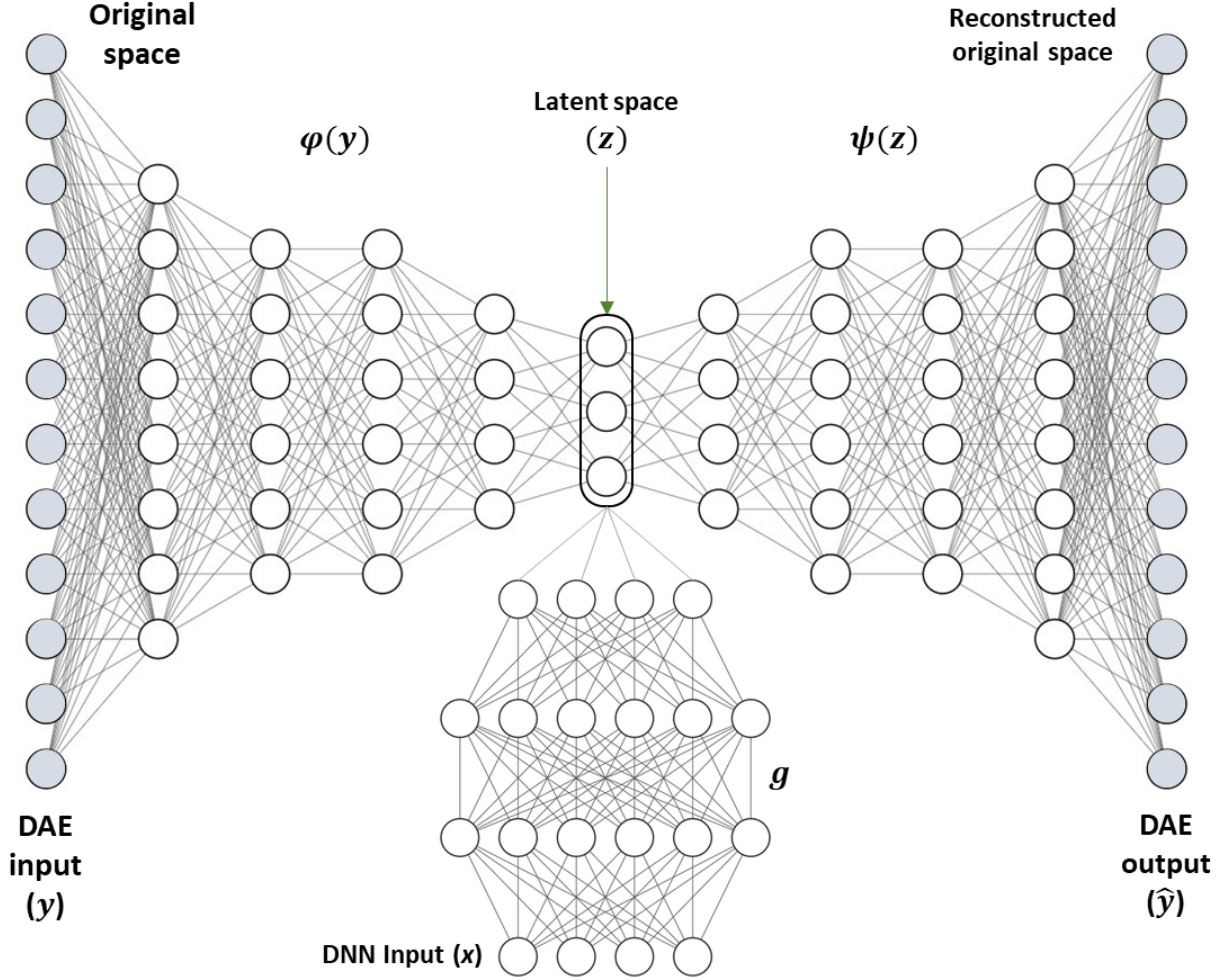

**Fig. 4.** Architecture of the proposed hybrid model

As indicated before, a weighted loss function is introduced to promote a balance between DAE and DNN training. Specifically, the total loss function of the hybrid model ($\mathcal{L}_{net}$) is decomposed into three elements, namely the DAE loss function ($\mathcal{L}_r$), the DNN loss function ($\mathcal{L}_z$), and the DNN-decoder loss function ($\mathcal{L}_x$). The total loss function can be then expressed as:

$$\mathcal{L}_{net} = \lambda_1 \mathcal{L}_r + \lambda_2 \mathcal{L}_z + \lambda_3 \mathcal{L}_x \tag{3}$$

where $\lambda_1$, $\lambda_2$, and $\lambda_3$ = weight parameters of the loss function, considered as system hyperparameters. The first loss function $\mathcal{L}_r$ is related to the DAE model and ensures that it reconstructs successfully the high-dimensional vector $\boldsymbol{y}$ given a predefined architecture which contains the low-dimensional latent space $\boldsymbol{z}$. The $\mathcal{L}_r$ can be expressed as:

$$\mathcal{L}_r = ||y - \psi(z)||_2^2 \tag{4}$$

where $||.||_2^2$ = the $L_2$ norm. The second loss function $\mathcal{L}_z$ is related to the regression model (i.e., DNN model), therefore it ensures that the DNN model accurately predicts the latent outputs $\boldsymbol{z}$ generated by the autoencoder, given the input vector $\boldsymbol{x}$ (i.e., storm parameters). The $\mathcal{L}_z$ loss function can be expressed as:

$$\mathcal{L}_z = ||z - g(x)||_2^2 \tag{5}$$

where g = regression function which is represented here by the DNN model. The third loss function $\mathcal{L}_x$ which represents the DNN-decoder loss, ensures that the transformation of the DNN predicted values, through the decoder, are consistent with the actual values in the original high-dimensional space. The DNN-decoder loss function can be expressed as:

$$\mathcal{L}_x = ||y - \psi(g(x))||_2^2 \tag{6}$$

Finally, once the model is trained, the predictive model, as shown in Fig. 5, can efficiently predict the high-dimensional vector (i.e., peak storm surge/significant wave height) based on any given input scenario $\boldsymbol{x}$ (i.e., storm parameters). Therefore, only the DNN, which predicts the latent output vectors, coupled with the decoder, which predicts the high-dimensional output vector over the entire region, are required.

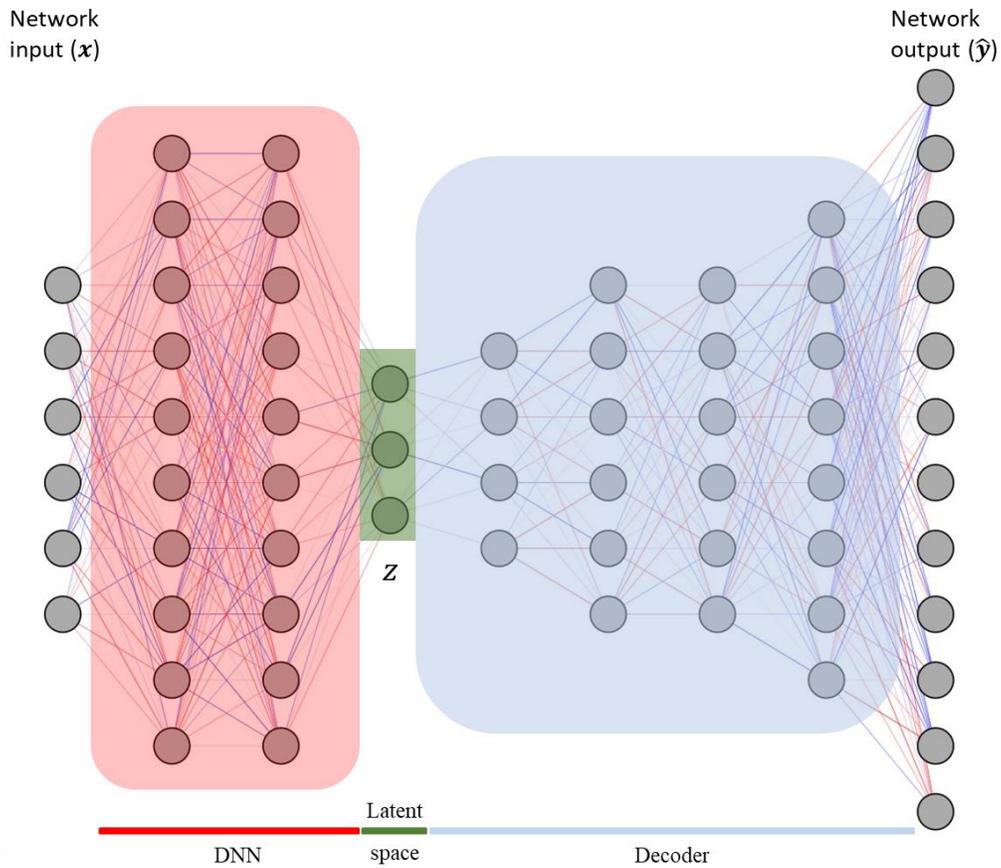

**Fig. 5.** Architecture of the predictive model

## 3. Case Study

In this study, several locations along New York (NY) and New Jersey (NJ) states were selected to simulate both hurricane-induced storm surge and significant wave height. The coastal regions of NY and NJ are prone to hurricane-induced hazards especially storm surge and flooding since they have several low-lying areas. These locations are expected to experience more devastating damage from hurricanes due to the effects of climate change and sea-level rise (Colle et al., 2010; Wang et al., 2014). Hurricane Sandy is an example of a storm that hit several locations in NY and NJ leading to unprecedented damage due to storm surge and flooding. It caused 72 fatalities and more than $50 billion in losses by damaging critical infrastructures (Blake et al., 2013). Therefore, the rapid prediction of hurricane-induced hazards in these vulnerable regions is necessary to advance hurricane risk assessment, management, mitigation, and improve our preparedness and climate resilience (Snaiki and Parida, 2023a, 2023b).

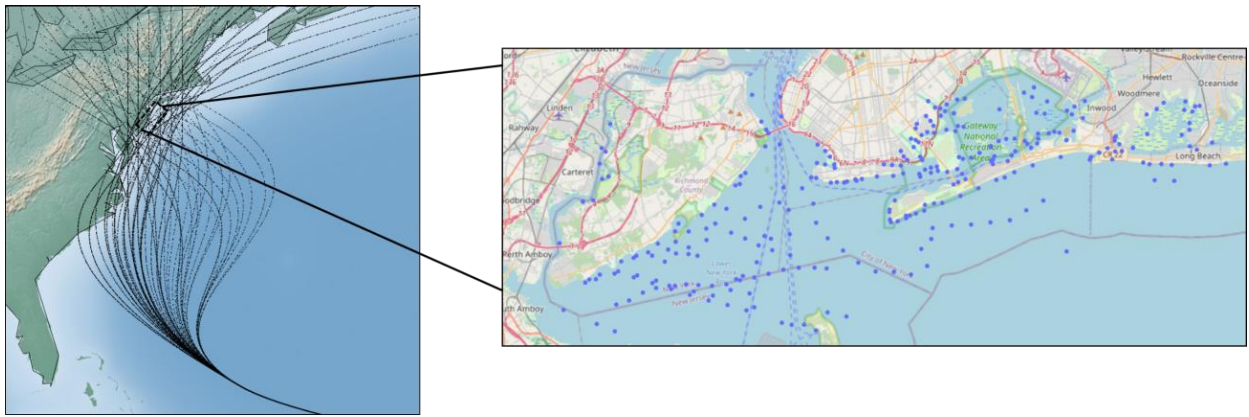

**Fig. 6.** Example of the selected NACCS synthetic tropical cyclones (left) and study area with the representation of save points (right).

Both storm surge and significant wave height will be predicted based on the proposed hybrid model which consists of a coupled DAE and DNN models trained simultaneously. To highlight the superior performance of the proposed model, it will be compared with a decoupled DAE and DNN model and another model in which the dimensionality reduction algorithm is carried out using the PCA technique along with a DNN model.

### 3.1. Database

The training/testing data are retrieved from the North Atlantic Comprehensive Coastal Study (NACCS) database which contains high-fidelity simulations corresponding to 1050 synthetic tropical cyclones. This large database is obtained from the Coastal Hazards System (CHS) v2.0 web-tool developed by the U.S. Army Corps of Engineers (USACE) which provides both storm-induced surge and significant wave height (Nadal-Caraballo et al., 2020). Specifically, the 'Base Conditions' were selected from the NACCS study (therefore, no tides are involved in the simulations). With this scenario and the selected save points, as geographically represented in Fig. 6, the peak storm surge and wave height have been retrieved from the database. In addition, the reference locations (in terms of latitude and longitude for both landfalling and bypassing storms) and the corresponding storm parameters related to the peak storm surge and wave height were also downloaded from the NACCS database. The storm surge and significant wave height responses are simulated using the ADvanced CIRCulation (ADCIRC) and the Steady State Spectral WAVE (STWAVE) models, respectively (Luettich et al., 1992; Smith et al., 1999; Cialone et al., 2015). The data preprocessing step indicated that out of the 1050 storms, only 1031 synthetic storms were successfully recorded. A total of 289 coastal save points are selected in this study area, which are distributed along critical regions of NY and NJ as indicated in Fig. 6. It should be noted that the 289 points were randomly selected; therefore, the proposed methodology can be readily applied for other save points. The database provides the peak storm surge and significant wave height values (output vector) along with six hurricane parameters (input vector, Eq. 1). It should be mentioned that the significant wave height ($H_s$) values were not reported for 31 save points out of the 289 selected points, therefore, they were not included in the simulation of $H_s$.

## 3.2. Training performance

In this study, two models were developed corresponding to the prediction of the peak storm surge and significant wave height, respectively. The dataset (i.e., the storm scenarios represented by the six storm parameters with their corresponding peak surge [or significant wave height] values over all save points) was divided randomly into training (70%), validation (15%), and testing (15%) sets. In addition, only the storms entering a 250-km-radius subregion centered on the studied area were retained which has led to a total of 414 synthetic storms. The storms located outside this region have little or no effects on the storm surge and significant wave height values.

As indicated in section 2.3, the loss function of the proposed DAE-DNN model is a weighted sum of three components, namely, the autoencoder loss function $\mathcal{L}_r$ (Eq. 4), the deep neural network's loss function $\mathcal{L}_z$ (Eq. 5), and the reconstruction loss function $\mathcal{L}_x$ (Eq. 6). Training the proposed hybrid model implies the minimization of the weighted loss function. To achieve this goal, several tuning parameters need to be carefully selected. This includes the network architecture (e.g., number of hidden layers for both DNN and DAE, activation function, number of neurons per layer, and learning rate) along with the weight parameters of the loss function (i.e., $\lambda_1$, $\lambda_2$, and $\lambda_3$). The trial-and-error technique is utilized in this study to find the best tuning parameters. The Xavier initialization technique is used for the weight initialization of the hybrid model and a Relu activation function is selected for both DNN and DAE models (except for the last layer where a linear function is used instead). In addition, the bias vectors are initialized to 0. Adam optimizer and a value of 0.001 for the learning rate are selected for both storm surge and significant wave height models. The training of the former required almost 6000 epochs, while it required almost 8000 epochs for the latter. One of the key hyperparameters is the number of latent neurons. To identify this number, a step-by-step approach is followed here (Champion et al., 2019). First, the minimum number of latent neurons is determined based on a typical autoencoder model (without coupling with a DNN model). This number is then used for the proposed hybrid model and increased until an appropriate model is identified.

The performance of the proposed hybrid DAE-DNN model is assessed through the mean squared error (MSE) and R-squared parameter ($R^2$) which are defined as follows:

$$MSE = \frac{1}{N}\sum_{i=1}^{N}(y_i - \hat{y}_i)^2 \tag{8}$$

$$R^2 = 1 - \frac{\sum_{i=1}^{N}(y_i - \hat{y}_i)^2}{\sum_{i=1}^{N}(y_i - \bar{y})^2} \tag{9}$$

where $N$ = number of outputs; $y$ = actual values given to the model; $\hat{y}$ = predicted values by the model; and $\bar{y}$ = mean of the actual (true) values. The results for training/validation for both storm surge and significant wave height are presented in the next section.

### 3.2.1. Storm surge

The deep autoencoder architecture consists of an input layer of 289 nodes (corresponding to the number of save points), followed by 2 hidden layers (both layers have 128 neurons). This part is also denoted as the encoder. It is then followed by the latent space which has 4 neurons. The decoder also has two hidden layers (both layers have 128 neurons) and an output layer with 289 nodes. On the other hand, the DNN architecture consists of an input layer with 6 nodes corresponding to the storm parameters (e.g., central pressure deficit), followed by two hidden layers (both layers have 64 neurons). The selected loss function weights $\lambda_1$, $\lambda_2$, and $\lambda_3$ are 1, 0.6, and 0.6, respectively. Figure 7 depicts the performance of the model training/validation in terms of the selected loss functions. Specifically, four loss functions were presented, namely the total loss of the hybrid model $\mathcal{L}_{net}$ (Eq. 3), the DAE loss $\mathcal{L}_r$ (Eq. 4), the DNN loss $\mathcal{L}_z$ (Eq. 5), and the DNN-decoder loss $\mathcal{L}_x$ (Eq. 6). The training results indicate that the loss functions are decreasing with increasing numbers of epochs for both training and validation. The total MSE values are 0.02 m$^2$ and 0.06 m$^2$ for training and validation, respectively.

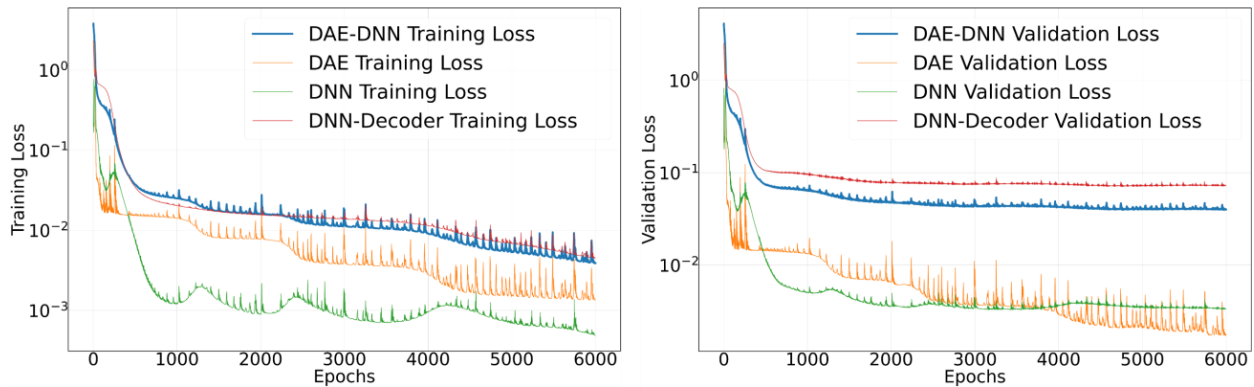

**Fig. 7.** Performance of the training process of the hybrid model for storm surge prediction for training (left) and validation (right)

In addition, the obtained R$^2$ values corresponding to the proposed hybrid model are 0.97 and 0.92 for training and validation, respectively, suggesting the good performance capacity of the model. To assess the performance of the proposed model, two additional models are implemented. The first model consists of a DAE and DNN which are trained separately (decoupled model). The second model is similar to the first one but PCA has been used for dimensionality reduction instead of the DAE algorithm. The simulation results are summarized in Table 1. Compared to the two decoupled models (i.e., decoupled DAE-DNN and decoupled PCA-DNN), the hybrid model has lower MSE values and higher R$^2$ values for both training and validation. For instance, the R$^2$ (MSE) value for training has reached the value of 0.97 (0.02 m$^2$) for the proposed model compared to 0.89 (0.07 m$^2$) and 0.92 (0.06 m$^2$) for the decoupled DAE-DNN and decoupled PCA-DNN, respectively. The comparison indicates clearly the superior prediction capabilities of the proposed model.

**Table 1.** Comparison of the performance of the training process of different models for storm surge prediction

| Models | MSE (m²) | | | R² score | | |
|---|---|---|---|---|---|---|
| | Training Set | Validation Set | Testing Set | Training Set | Validation Set | Testing Set |
| DAE-DNN | 0.02 | 0.06 | 0.07 | 0.97 | 0.92 | 0.91 |
| Decoupled DAE-DNN | 0.07 | 0.09 | 0.16 | 0.89 | 0.85 | 0.75 |
| Decoupled PCA-DNN | 0.06 | 0.08 | 0.12 | 0.92 | 0.90 | 0.85 |

### 3.2.2. Significant wave height

The deep autoencoder architecture of the proposed hybrid model contains an input layer of 258 nodes which correspond to the number of save points, followed by two hidden layers (the first one has 128 neurons and the second one has 64 neurons). The latent space has 6 neurons. The decoder part also has two hidden layers (the first one has 64 neurons and the second one has 128 neurons). Additionally, the architecture of the DNN model consists of an input layer with 6 nodes corresponding to the storm parameters, followed by two hidden layers (both layers have 64 neurons). The selected loss function weights $\lambda_1$, $\lambda_2$, and $\lambda_3$ are 1, 0.7, and 0.001, respectively. The model training and validation results are shown in Fig. 8. Similar to the surge simulation, the training results show that for both training and validation, the loss functions decrease with increasing numbers of epochs. The total MSE values are 0.02 m² and 0.03 m² for training and validation, respectively.

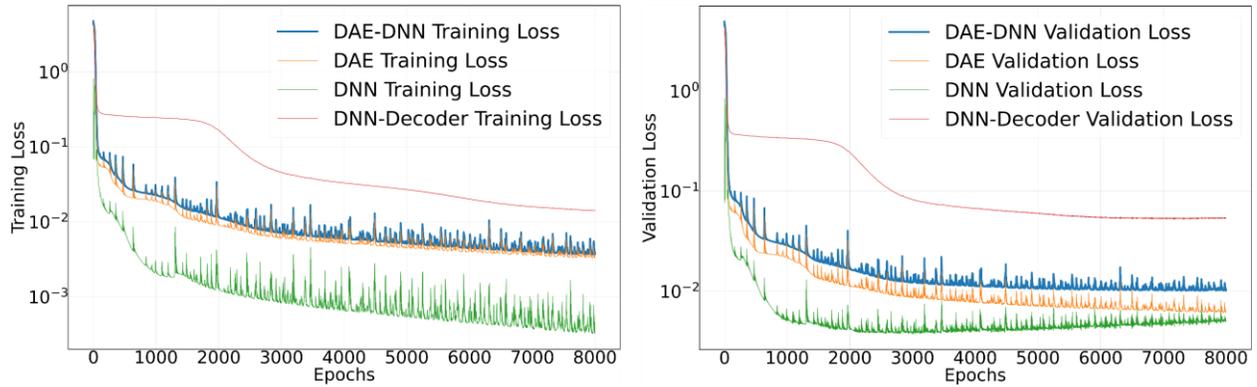

**Fig. 8.** Performance of the hybrid model for significant wave height prediction for training (left) and validation (right)

Additionally, the obtained $R^2$ values for the proposed hybrid model are 0.90 for training and 0.85 for validation. Two models, previously defined in Sect. 3.2.1, were also trained to verify the superior performance of the proposed hybrid model, namely the decoupled DAE-DNN and PCA-DNN. Table 2 provides a summary of the simulation results. The hybrid model clearly outperforms the two decoupled models for both training and validation since it has lower MSE values and higher $R^2$ values. For instance, the proposed model's $R^2$ (MSE) value for training has reached the value of 0.90 (0.02 m$^2$) compared to 0.84 (0.06 m$^2$) and 0.84 (0.08 m$^2$) for the decoupled DAE-DNN and PCA-DNN, respectively.

**Table 2.** Comparison of the performance of the training process of different models for significant wave height prediction

| Models | MSE (m$^2$) | | | R$^2$ score | | |
|---|---|---|---|---|---|---|
| | Training Set | Validation Set | Testing Set | Training Set | Validation Set | Testing Set |
| DAE-DNN | 0.02 | 0.03 | 0.04 | 0.90 | 0.85 | 0.82 |
| Decoupled DAE-DNN | 0.06 | 0.07 | 0.12 | 0.84 | 0.82 | 0.78 |
| Decoupled PCA-DNN | 0.08 | 0.12 | 0.16 | 0.84 | 0.78 | 0.72 |

Once the proposed model is trained, the latent space can be identified. Therefore, it is important to assess the performance of the predictive model, which is formed as the combination of the DNN and decoder models, as shown in Fig. 5, using the testing (15%) set. As mentioned earlier, the predictive model takes the six hurricane parameters as inputs (i.e., Eq. 1) and predicts the peak storm surge/significant wave height values over all save points. The testing results indicated a good performance for predicting the peak storm surge (significant wave height) with R$^2$ and MSE values of 91% (82%) and 0.07 m$^2$ (0.04 m$^2$), respectively. In the next section, two randomly selected storm scenarios from the testing set are chosen to visualize the peak storm surge and significant wave height values across all save points.

## 4. Application

Two case studies corresponding to storm surge and significant wave height prediction will be presented to highlight the accurate prediction capacities of the proposed hybrid model. The first scenario involves a relatively weak storm with a low-pressure deficit, while the second scenario features an intense storm with a high-pressure deficit. This selection enables an evaluation of the model's performance across a wide spectrum of storm intensities, demonstrating its broad applicability and accuracy. The storm parameters of the two selected scenarios are summarized in Table 3.

**Table 3.** Storm parameters for storm surge and significant wave height prediction

| Parameter | $C_p$(hPa) | $\theta$(°) | $R_{max}$(km) | $LAT$(°) | $LON$(°) | $V_f(km/h)$ |
|---|---|---|---|---|---|---|
| First scenario | 28 | -60 | 58 | 38.89 | -75.27 | 13 |
| Second scenario | 88 | -60 | 50 | 38.23 | -75.14 | 37 |

With the values of the storm parameters listed in Table 3 and the trained hybrid DAE-DNN model, the peak storm surge and significant wave height can be obtained. The simulation results for peak storm surge under two scenarios are illustrated in Fig 9.

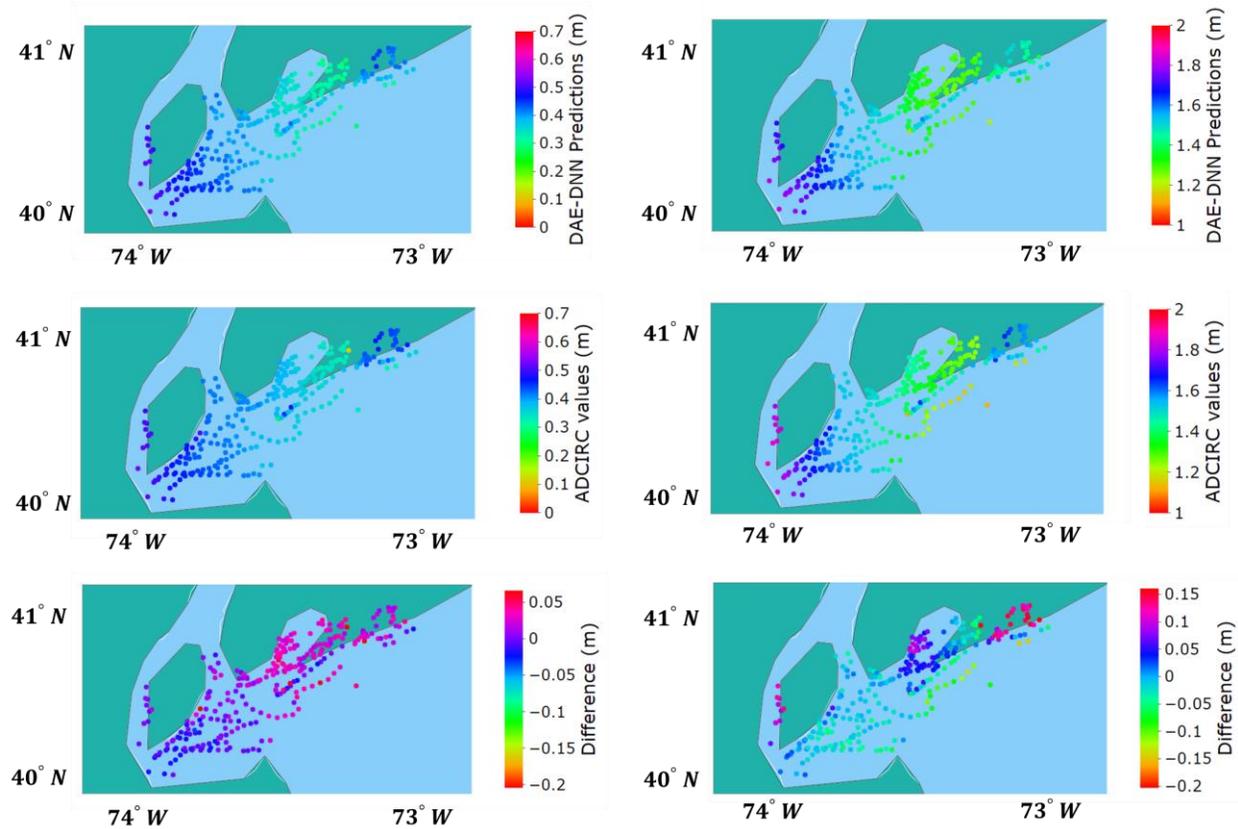

**Fig. 9.** ADCIRC-based and simulated-based peak storm surge of the first storm scenario (left column) and second storm scenario (right column)

As shown in Fig. 9, an excellent agreement between the simulated peak storm surge and ADCIRC-based simulation results is observed with a difference (i.e., ADCIRC-based results minus model predicted values) that is not exceeding 0.2 m for both scenarios. Similarly, with the same storm parameters listed in Table 3, the peak significant wave height was predicted based on the two selected scenarios. The simulation results are depicted in Fig 10.

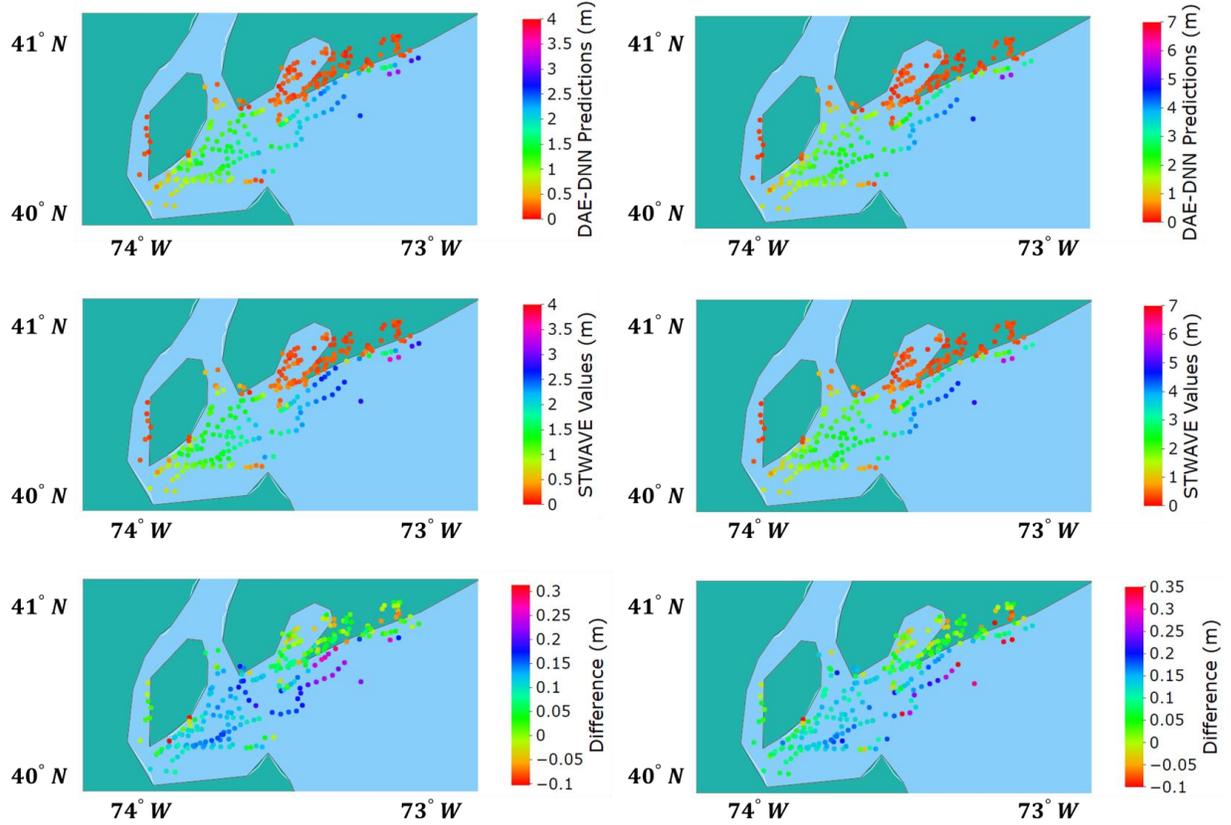

**Fig. 10.** STWAVE-based and simulated-based peak significant wave height of the first storm scenario (left column) and second storm scenario (right column)

Excellent agreement between the simulated peak significant wave height and STWAVE-based simulation results were achieved with a difference not exceeding 0.35 m for both scenarios.

A sensitivity analysis was also carried out in terms of the central pressure deficit parameter, which significantly affects the storm intensity and hence the storm surge and significant wave height, to examine the validity of the obtained results. In this case, the first storm scenario as shown in Table 3 was selected. It is shown that, based on the base scenario (1ˢᵗ scenario) of Table 3 with three different central pressure values $p_c = 983 \ hPa$ ($C_p \approx 30 \ hPa$), $p_c = 953 \ hPa$ ($C_p \approx 60 \ hPa$), and $p_c = 923 \ hPa$ ($C_p \approx 90 \ hPa$), the predicted storm surge and significant wave height are substantially altered as indicated in Fig. 11. Specifically, for $p_c = 983 \ hPa$, which corresponds to a category 1 hurricane, the storm surge values do not exceed 0.6 m compared to maximum values of 1.4 m for category 3 hurricane (i.e., $p_c = 953 \ hPa$). For the worst-case scenario (i.e., $p_c = 923 \ hPa$ which is nearly a category 5 hurricane), the maximum storm surge values can reach a 1.8 m height in few locations as indicated in Fig. 11. Similarly, the significant wave height reaches a maximum value of 5 m (6 m) in a few locations for $p_c = 983 \ hPa$ ($p_c = 953 \ hPa$), and increases to a maximum value of 7 m for the worst-case scenario (i.e., $p_c = 923 \ hPa$). Those results are consistent with the inherent physics of storm surge/waves since intense storms usually lead to significant values of storm surge/waves.

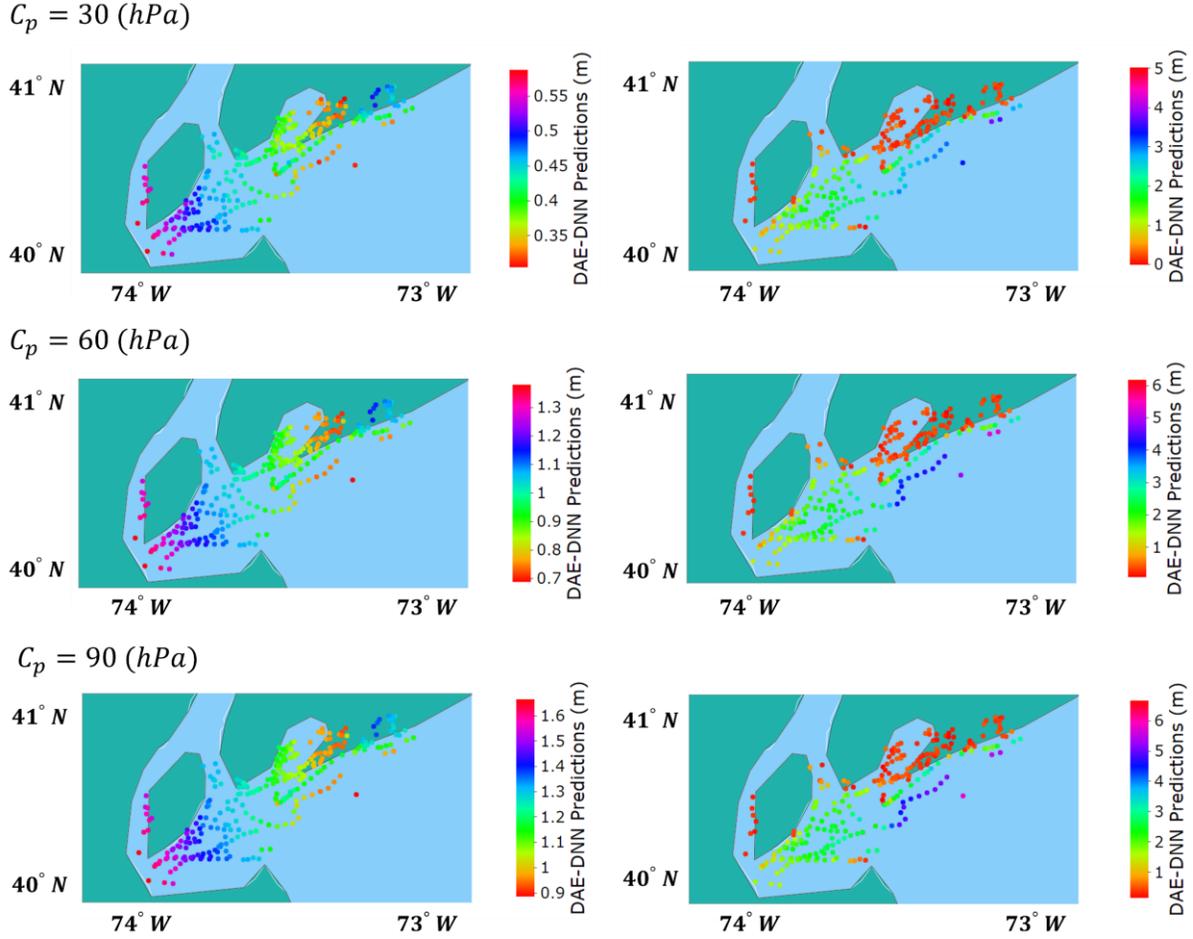

$C_p = 30\ (hPa)$

$C_p = 60\ (hPa)$

$C_p = 90\ (hPa)$

**Fig. 11.** Predicted values for storm surge (left) and significant wave height (right) given different values of central pressure

## 5. Discussion

To address the critical need for rapid and accurate prediction of storm surge and wave height induced by cyclones, this study proposes a novel hybrid model. High-fidelity numerical models, despite their accuracy, are computationally expensive, limiting their use in real-time applications. The proposed model combines the strengths of dimensionality reduction via deep autoencoders and data-driven mapping with deep neural networks. It efficiently captures a low-dimensional representation of the coastal system while simultaneously mapping storm parameters onto this latent space. This enables rapid assessment of peak storm surge and significant wave height over large coastal regions. Compared to standard decoupled approaches (Jia and Taflanidis, 2013; Atteia et. al., 2022), the hybrid model demonstrates superior performance, achieving high accuracy as highlighted in Fig. 7 and 8 as well as Table 1 and 2. Furthermore, its generalizability allows application to other geographical locations with necessary customization in terms of data and model structure. This methodology offers a promising solution for rapid and accurate prediction, contributing to coastal hazard mitigation and preparedness efforts.

While the proposed hybrid model exhibits impressive accuracy in peak surge and wave simulations, some limitations need to be addressed. For example, the identification of the model hyperparameters is based on a trial-and-error approach. Despite achieving good performance, this approach is inefficient and may not identify the optimal setup. Employing Bayesian optimization to explore the hyperparameter space could lead to improved configurations and training efficiency. Notably, the number of intrinsic coordinates in the autoencoder's latent space significantly impacts interpretation and the associated dynamical model. In this study, identification of the minimum number required a two-step process: first, determining it for a standard autoencoder, then adapting it for the hybrid model. This approach, though effective, is tedious and does not guarantee optimal configuration. To address this, developing a hybrid model equipped with a suitable optimization technique to automatically select the optimal number of latent neurons would enhance the simulation results. To further enhance the model, incorporating the effects of astronomical tides is recommended. While a linear addition to the model's output is feasible, it may not fully capture the potential nonlinear effects of the tides (Xiao et al., 2021). Therefore, incorporating the tides as input to the DNN model is likely to result in a more accurate representation. In addition, expanding the model to simulate time series of both surge and wave height (Saviz Naeini & Snaiki, 2024) necessitates further adaptation. Replacing DNN with advanced surrogate models like Long Short-Term Memory (LSTM) and employing other dimensionality reduction techniques (e.g., convolutional autoencoders) capable of capturing spatiotemporal patterns within data are essential. Additionally, exploring techniques such as knowledge-enhanced neural networks (Karniadakis et al., 2021; Saviz Naeini & Snaiki, 2024; Snaiki & Wu, 2019, 2022) could circumvent the need for larger databases by leveraging both data and underlying physical principles.

## 6. Conclusion

In this study, a novel hybrid machine learning model has been proposed for rapid prediction of peak storm surge and waves over an extended coastal region for both landfalling and bypassing storms. The proposed hybrid model trains simultaneously a deep autoencoder (DAE) and a deep neural network (DNN) based on a unique weighted loss function. While the DAE identifies a low-dimensional representation of the high-dimensional spatial system, the DNN maps the storm parameters (e.g., central pressure deficit and radius of maximum wind) to the obtained low-dimensional latent space. To demonstrate the superior performance of the hybrid model, the peak storm surge and significant wave height were predicted over several coastal locations within NY and NJ. A total of 289 coastal locations were selected which serve as the input to the DAE model. In addition, six storm parameters have been selected as input to the DNN model, namely reference latitude, reference longitude, heading angle, central pressure deficit, translation speed, and radius of maximum winds. The proposed technique was further compared with two decoupled models consisting of a dimensionality reduction technique (PCA and DAE) and a regression model based on DNN which are trained separately. The hybrid model outperformed the decoupled PCA-DNN and DAE-DNN. For example, the $R^2$ (MSE) for training of the storm surge-based model was 0.97 (0.02 $m^2$), 0.89 (0.07 $m^2$), 0.92 (0.06 $m^2$) for the hybrid model, decoupled DAE-DNN, and decoupled PCA-DNN, respectively. Similarly, the $R^2$ (MSE) for training of the significant wave height-based model was 0.90 (0.02 $m^2$), 0.84 (0.06 $m^2$), 0.84 (0.08 $m^2$) for the hybrid model, decoupled DAE-DNN, and decoupled PCA-DNN, respectively. In addition, the testing results corresponding to the predictive model indicated a good performance for predicting the peak storm surge (significant wave height) with $R^2$ and MSE values of 91% (82%) and 0.07 $m^2$ (0.04 $m^2$), respectively. Consequently, high accuracy and computational efficiency are observed for the hybrid model which could be readily integrated as part of an early warning system or used for probabilistic risk assessment, and rapid prediction of waves and storm surge.

**Declaration of Competing Interest**

The authors declare that they have no known competing financial interests or personal relationships that could have appeared to influence the work reported in this paper.

**Acknowledgements**

This work was supported by the Natural Sciences and Engineering Research Council of Canada (NSERC) [grant number CRSNG RGPIN 2022-03492]

**Appendix A**

The neural network algorithms were first inspired by biological neural networks to mimic human brain neural activity (Liu et al., 2021). Neural networks are typically composed of an input layer, several hidden layers, and an output layer as shown in Fig. 1. Each hidden layer consists of several neurons (Leung et al., 2003). The neurons are considered as a computational unit that receive one or more inputs and produce an output based on a specific activation function. The output of each neuron can be then obtained as follows:

$$c = \Phi(\sum_{i=1}^{n} w_i a_i + b) \tag{10}$$

where $\Phi$ = activation function which is a mathematical function that introduces nonlinear characteristics to the neuron's output; $b$ = bias; and $w_i$ = weight associated with input $x_i$. The activation function captures the nonlinearities within the data and can be selected from well-known functions including Hyperbolic-tangent, Sigmoid, Relu, among others (Zhang et al., 2018). In order to train a typical ANN model, its architecture should be initially defined. The weights and biases should also be initialized. Several methods are being used for weight initialization such as the random initialization, Xavier initialization, orthogonal initialization, etc. On the other hand, the biases are usually set to zero to make the training process simpler. The weights of the inputs are adjusted during the training process of the neural network in order to optimize the performance of the network (Larochelle et al.,2009). Specifically, the training process of typical ANN models consists of two major steps. In the first step, also denoted as the feedforward step, the ANN output is computed as a result of the aggregation of the neurons output (Eq. 10). In the second step, also denoted as the backpropagation step, the derived output is compared to the target value, and the obtained errors are subsequently backpropagated through the network. The best weights and biases are found by minimizing the loss function using an optimizer algorithm (e.g., gradient-descent, stochastic gradient-descent or Adam) while repeating the feedforward and backpropagation steps for a certain number of epochs (the training can also stop once a given convergence criterion is met). The learning rate, an important hyperparameter in the ANN optimization, determines the size of the step taken during each training iteration (Rumelhart et al., 1986; LeCun et al., 1989; Ruder, 2016). Its magnitude affects the learning ability of the model based on the training data, with higher values resulting in faster convergence but a higher risk of overshooting the global minimum. On the other hand, lower values will lead to slower convergence due to smaller steps, but with the potential for better accuracy. The trained model should be evaluated on a test set to assess its performance. The performance of the ANN models depends on the selected hyperparameters including the number of inputs, number of hidden layers, number of neurons in each layer, weight initialization, activation function, learning rate, and the optimization algorithm (Bardenet et al., 2013).

## Data availability

Data will be made available on request.

## References

Abdi, H., & Williams, L. J. (2010). Principal component analysis. *Wiley interdisciplinary reviews: computational statistics*, *2*(4), 433-459.

Adeli, E., Sun, L., Wang, J., & Taflanidis, A. A. (2022). An advanced spatio-temporal convolutional recurrent neural network for storm surge predictions. *arXiv preprint arXiv:2204.09501*.

Al Kajbaf, A., & Bensi, M. (2020). Application of surrogate models in estimation of storm surge: A comparative assessment. *Applied Soft Computing*, *91*, 106184.

Atteia, G., Collins, M. J., Algarni, A. D., & Samee, N. A. (2022). Deep-Learning-Based Feature Extraction Approach for Significant Wave Height Prediction in SAR Mode Altimeter Data. *Remote Sensing*, *14*(21), 5569.

Bai, L. H., & Xu, H. (2022). Accurate storm surge forecasting using the encoder–decoder long short term memory recurrent neural network. *Physics of Fluids*, *34*(1), 016601.

Bajo, M., & Umgiesser, G. (2010). Storm surge forecast through a combination of dynamic and neural network models. *Ocean Modelling*, *33*(1-2), 1-9.

Bardenet, R., Brendel, M., Kégl, B., & Sebag, M. (2013, May). Collaborative hyperparameter tuning. In *International conference on machine learning* (pp. 199-207). PMLR.

Bass, B., & Bedient, P. (2018). Surrogate modeling of joint flood risk across coastal watersheds. *Journal of Hydrology*, *558*, 159–173.

Berbić, J., Ocvirk, E., Carević, D., & Lončar, G. (2017). Application of neural networks and support vector machine for significant wave height prediction. *Oceanologia*, *59*(3), 331–349.

Bezuglov, A., Blanton, B., & Santiago, R. (2016). Multi-output artificial neural network for storm surge prediction in north carolina. *arXiv preprint arXiv:1609.07378*.

Blake, E. S., Kimberlain, T. B., Berg, R. J., Cangialosi, J. P., & Beven Ii, J. L. (2013). Tropical cyclone report: Hurricane sandy. *National Hurricane Center*, *12*, 1–10.

Booij, N., Ris, R. C., & Holthuijsen, L. H. (1999). A third-generation wave model for coastal regions: 1. Model description and validation. *Journal of Geophysical Research: Oceans*, *104*(C4), 7649–7666.

Bretschneider, C. L. (1967). Storm surges. In *Advances in hydroscience* (Vol. 4, pp. 341-418). Elsevier.

Callens, A., Morichon, D., Abadie, S., Delpey, M., & Liquet, B. (2020). Using Random forest and Gradient boosting trees to improve wave forecast at a specific location. *Applied Ocean Research*, *104*, 102339.

Champion, K., Lusch, B., Kutz, J. N., & Brunton, S. L. (2019). Data-driven discovery of coordinates and governing equations. *Proceedings of the National Academy of Sciences*, *116*(45), 22445-22451.

Chen, Q., Wang, L., & Tawes, R. (2008). Hydrodynamic Response of Northeastern Gulf of Mexico to Hurricanes. *Estuaries and Coasts*, *31*(6), 1098–1116.

Chen, W.-B., Liu, W.-C., & Hsu, M.-H. (2012). Predicting typhoon-induced storm surge tide with a two-dimensional hydrodynamic model and artificial neural network model. *Natural Hazards and Earth System Sciences*, *12*(12), 3799–3809.

Cialone, M. A., Massey, T. C., Anderson, M. E., Grzegorzewski, A. S., Jensen, R. E., Cialone, A., Mark, D. J., Pevey, K. C., Gunkel, B. L., & McAlpin, T. O. (2015). *North Atlantic Coast Comprehensive Study (NACCS) coastal storm model simulations: Waves and water levels*.

Colle, B. A., Rojowsky, K., & Buonaito, F. (2010). New York City storm surges: Climatology and an analysis of the wind and cyclone evolution. *Journal of Applied Meteorology and Climatology*, *49*(1), 85-100.

Dinan, T. (2016). Potential increases in hurricane damage in the united states: implications for the federal budget. Congress of the United States, Congressional Budget Office.

Fan, S., Xiao, N., & Dong, S. (2020). A novel model to predict significant wave height based on long short-term memory network. *Ocean Engineering*, *205*, 107298.

Fan, Y., Ginis, I., & Hara, T. (2009). The Effect of Wind–Wave–Current Interaction on Air–Sea Momentum Fluxes and Ocean Response in Tropical Cyclones. *Journal of Physical Oceanography*, *39*(4), 1019–1034.

French, J., Mawdsley, R., Fujiyama, T., & Achuthan, K. (2017). Combining machine learning with computational hydrodynamics for prediction of tidal surge inundation at estuarine ports. *Procedia IUTAM*, *25*, 28–35.

Gao, R., Li, R., Hu, M., Suganthan, P. N., & Yuen, K. F. (2023). Significant wave height forecasting using hybrid ensemble deep randomized networks with neurons pruning. *Engineering Applications of Artificial Intelligence*, *117*, 105535.

Grossberg, S. (1988). Nonlinear neural networks: Principles, mechanisms, and architectures. *Neural Networks*, *1*(1), 17–61.

Hanson, J. L., Forte, M. F., Blanton, B., Gravens, M., & Vickery, P. (2013). Coastal Storm Surge Analysis: Storm Surge Results. *Report 5: Intermediate Submission No*, *3*.

Hashemi, M. R., Spaulding, M. L., Shaw, A., Farhadi, H., & Lewis, M. (2016). An efficient artificial intelligence model for prediction of tropical storm surge. *Natural Hazards*, *82*(1), 471–491.

Igarashi, Y., & Tajima, Y. (2021). Application of recurrent neural network for prediction of the time-varying storm surge. *Coastal Engineering Journal*, *63*(1), 68–82.

Irish, J. L., & Resio, D. T. (2010). A hydrodynamics-based surge scale for hurricanes. *Ocean Engineering*, *37*(1), 69–81.

Irish, J. L., Resio, D. T., & Cialone, M. A. (2009). A surge response function approach to coastal hazard assessment. Part 2: Quantification of spatial attributes of response functions. *Natural Hazards*, *51*(1), 183–205.

Jelesnianski, C. P. (1992). *SLOSH: Sea, lake, and overland surges from hurricanes* (Vol. 48). US Department of Commerce, National Oceanic and Atmospheric Administration, National Weather Service.

Jia, G., & Taflanidis, A. A. (2013). Kriging metamodeling for approximation of high-dimensional wave and surge responses in real-time storm/hurricane risk assessment. *Computer Methods in Applied Mechanics and Engineering*, *261–262*, 24–38.

Jia, G., Taflanidis, A. A., Nadal-Caraballo, N. C., Melby, J. A., Kennedy, A. B., & Smith, J. M. (2016). Surrogate modeling for peak or time-dependent storm surge prediction over an extended coastal region using an existing database of synthetic storms. *Natural Hazards*, *81*(2), 909–938.

Karniadakis, G.E., Kevrekidis, I.G., Lu, L., Perdikaris, P., Wang, S. and Yang, L., 2021. Physics-informed machine learning. *Nature Reviews Physics*, *3*(6), pp.422-440.

Kijewski-Correa, T., Taflanidis, A., Vardeman, C., Sweet, J., Zhang, J., Snaiki, R., ... & Kennedy, A. (2020). Geospatial environments for hurricane risk assessment: applications to situational awareness and resilience planning in New Jersey. *Frontiers in Built Environment*, *6*, 549106.

Kim, S.-W., Melby, J. A., Nadal-Caraballo, N. C., & Ratcliff, J. (2015). A time-dependent surrogate model for storm surge prediction based on an artificial neural network using high-fidelity synthetic hurricane modeling. *Natural Hazards*, *76*(1), 565–585.

Kyprioti, A. P., Taflanidis, A. A., Nadal-Caraballo, N. C., Yawn, M. C., & Aucoin, L. A. (2022). Integration of Node Classification in Storm Surge Surrogate Modeling. *Journal of Marine Science and Engineering*, *10*(4), 551.

Kyprioti, A. P., Taflanidis, A. A., Plumlee, M., Asher, T. G., Spiller, E., Luettich, R. A., Blanton, B., Kijewski-Correa, T. L., Kennedy, A., & Schmied, L. (2021). Improvements in storm surge surrogate modeling for synthetic storm parameterization, node condition classification and implementation to small size databases. *Natural Hazards*, *109*(2), 1349–1386.

Larochelle, H., Bengio, Y., Louradour, J., & Lamblin, P. (2009). Exploring strategies for training deep neural networks. *Journal of machine learning research*, *10*(1).

LeCun, Y., Boser, B., Denker, J. S., Henderson, D., Howard, R. E., Hubbard, W., & Jackel, L. D. (1989). Backpropagation Applied to Handwritten Zip Code Recognition. *Neural Computation*, *1*(4), 541–551.

Lee, J.-W., Irish, J. L., Bensi, M. T., & Marcy, D. C. (2021). Rapid prediction of peak storm surge from tropical cyclone track time series using machine learning. *Coastal Engineering*, *170*, 104024.

Lee, T.-L. (2006). Neural network prediction of a storm surge. *Ocean Engineering*, *33*(3–4), 483–494.

Leung, F. H. F., Lam, H. K., Ling, S. H., & Tam, P. K. S. (2003). Tuning of the structure and parameters of a neural network using an improved genetic algorithm. *IEEE Transactions on Neural Networks*, *14*(1), 79–88.

Lin, N., & Chavas, D. (2012). On hurricane parametric wind and applications in storm surge modeling. *Journal of Geophysical Research: Atmospheres*, *117*(D9).

Lin, N., Emanuel, K. A., Smith, J. A., & Vanmarcke, E. (2010). Risk assessment of hurricane storm surge for New York City. *Journal of Geophysical Research: Atmospheres*, *115*(D18).

Liou, C.-Y., Cheng, W.-C., Liou, J.-W., & Liou, D.-R. (2014). Autoencoder for words. *Neurocomputing*, *139*, 84–96.

Liu, C., Arnon, T., Lazarus, C., Strong, C., Barrett, C., & Kochenderfer, M. J. (2021). Algorithms for Verifying Deep Neural Networks. *Foundations and Trends® in Optimization*, *4*(3–4), 244–404.

Lockwood, J. W., Lin, N., Oppenheimer, M., & Lai, C. Y. (2022). Using Neural Networks to Predict Hurricane Storm Surge and to Assess the Sensitivity of Surge to Storm Characteristics. *Journal of Geophysical Research: Atmospheres*, *127*(24), e2022JD037617.

Luettich, R. A. (Richard A., Westerink, J. J., & Scheffner, N. W. (1992). *ADCIRC : an advanced three-dimensional circulation model for shelves, coasts, and estuaries. Report 1, Theory and methodology of ADCIRC-2DD1 and ADCIRC-3DL.*

Luettich, R. A., & Westerink, J. J. (2004). *Formulation and numerical implementation of the 2D/3D ADCIRC finite element model version 44. XX* (Vol. 20, pp. 74-74). Chapel Hill, NC, USA: R. Luettich.

Meng, F., Song, T., Xu, D., Xie, P., & Li, Y. (2021). Forecasting tropical cyclones wave height using bidirectional gated recurrent unit. *Ocean Engineering*, *234*, 108795.

Nadal-Caraballo, N. C., Campbell, M. O., Gonzalez, V. M., Torres, M. J., Melby, J. A., & Taflanidis, A. A. (2020). Coastal hazards system: a probabilistic coastal hazard analysis framework. *Journal of Coastal Research*, *95*(SI), 1211-1216.

Naeini, S.S. and Snaiki, R., 2024. A physics-informed machine learning model for time-dependent wave runup prediction. *Ocean Engineering*, *295*, p.116986.

Plumlee, M., Asher, T. G., Chang, W., & Bilskie, M. V. (2021). High-fidelity hurricane surge forecasting using emulation and sequential experiments.

Portnova-Fahreeva, A. A., Rizzoglio, F., Nisky, I., Casadio, M., Mussa-Ivaldi, F. A., & Rombokas, E. (2020). Linear and Non-linear Dimensionality-Reduction Techniques on Full Hand Kinematics. *Frontiers in Bioengineering and Biotechnology*, *8*, 429.

Ramos-Valle, A. N., Curchitser, E. N., Bruyère, C. L., & McOwen, S. (2021). Implementation of an Artificial Neural Network for Storm Surge Forecasting. *Journal of Geophysical Research: Atmospheres*, *126*(13), e2020JD033266.

Rao, N. B., & Mazumdar, S. (1966). A technique for forecasting storm waves. *MAUSAM*, *17*(3), 333-346.

Ruder, S. (2016). An overview of gradient descent optimization algorithms. *arXiv preprint arXiv:1609.04747*.

Rumelhart, D. E., Hinton, G. E., & Williams, R. J. (1986). Learning representations by back-propagating errors. *Nature 1986 323:6088*, *323*(6088), 533–536.

Saviz, S., & Snaiki, R. (2022). *Machine Learning Approximation for Rapid Prediction of High-Dimensional Storm Surge and Wave Responses*. Proceedings of the Canadian Society of Civil Engineering Annual Conference 2022, Whistler, BC, Canada.

Smith, J. M., Resio, D. T., & Zundel, A. K. (1999). *STWAVE: Steady-state spectral wave model. Report 1. User's Manual For STWAVE Version 2.0.* ARMY ENGINEER WATERWAYS EXPERIMENT STATION VICKSBURG MS COASTAL AND HYDRAULICS LAB.

Snaiki, R., & Parida, S. S. (2023a). A data-driven physics-informed stochastic framework for hurricane-induced risk estimation of transmission tower-line systems under a changing climate. *Engineering Structures*, *280*, 115673.

Snaiki, R., & Parida, S. S. (2023b). Climate change effects on loss assessment and mitigation of residential buildings due to hurricane wind. *Journal of Building Engineering*, *69*, 106256.

Snaiki, R., & Wu, T. (2019). Knowledge-enhanced deep learning for simulation of tropical cyclone boundary-layer winds. *Journal of Wind Engineering and Industrial Aerodynamics*, *194*, 103983.

Snaiki, R., Wu, T., Whittaker, A. S., & Atkinson, J. F. (2020). Hurricane wind and storm surge effects on coastal bridges under a changing climate. *Transportation research record*, *2674*(6), 23-32.

Snaiki, R., & Wu, T. (2022). Knowledge-enhanced deep learning for simulation of extratropical cyclone wind risk. *Atmosphere*, *13*(5), 757.

Song, T., Han, R., Meng, F., Wang, J., Wei, W., & Peng, S. (2022). A Significant Wave Height Prediction Method Based on Deep Learning Combining the Correlation between Wind and Wind Waves. *Frontiers in Marine Science*, 1931.

Sze, V., Chen, Y. H., Yang, T. J., & Emer, J. S. (2017). Efficient Processing of Deep Neural Networks: A Tutorial and Survey. *Proceedings of the IEEE*, *105*(12), 2295–2329.

Sztobryn, M. (2003). Forecast of storm surge by means of artificial neural network. *Journal of Sea Research*, *49*(4), 317–322.

Tadesse, M., Wahl, T., & Cid, A. (2020). Data-Driven Modeling of Global Storm Surges. *Frontiers in Marine Science*, *7*, 260.

Taflanidis, A. A., Jia, G., Kennedy, A. B., & Smith, J. M. (2013). Implementation/optimization of moving least squares response surfaces for approximation of hurricane/storm surge and wave responses. *Natural Hazards: Journal of the International Society for the Prevention and Mitigation of Natural Hazards*, *66*(2), 955–983.

Thomas, T. J., & Dwarakish, G. S. (2015). Numerical Wave Modelling – A Review. *Aquatic Procedia*, *4*, 443–448.

Van Der Maaten, L., Postma, E., & Van den Herik, J. (2009). Dimensionality reduction: a comparative review. *J Mach Learn Res*, *10*(66–71), 13.

Wamsley, T. V., Cialone, M. A., Smith, J. M., Ebersole, B. A., & Grzegorzewski, A. S. (2009). Influence of landscape restoration and degradation on storm surge and waves in southern Louisiana. *Natural Hazards*, *51*(1), 207–224.

Wang, H. V., Loftis, J. D., Liu, Z., Forrest, D., & Zhang, J. (2014). The storm surge and sub-grid inundation modeling in New York City during Hurricane Sandy. *Journal of Marine Science and Engineering*, *2*(1), 226-246.

Wetzel, S. J. (2017). Unsupervised learning of phase transitions: From principal component analysis to variational autoencoders. *Physical Review E*, *96*(2), 022140.

Wu, T., & Snaiki, R. (2022). Applications of machine learning to wind engineering. *Frontiers in Built Environment*, *8*, 811460.

Xiao, Z., Yang, Z., Wang, T., Sun, N., Wigmosta, M., & Judi, D. (2021). Characterizing the non-linear interactions between tide, storm surge, and river flow in the delaware bay estuary, United States. *Frontiers in Marine Science*, *8*, 715557.

Zhang, H., Weng, T. W., Chen, P. Y., Hsieh, C. J., & Daniel, L. (2018). Efficient neural network robustness certification with general activation functions. *Advances in neural information processing systems*, *31*.

Zhang, J., Taflanidis, A. A., Nadal-Caraballo, N. C., Melby, J. A., & Diop, F. (2018). Advances in surrogate modeling for storm surge prediction: storm selection and addressing characteristics related to climate change. *Natural Hazards*, *94*(3), 1225–1253.

Zhang, K., Douglas, B. C., & Leatherman, S. P. (2000). Twentieth-Century Storm Activity along the U.S. East Coast. *Journal of Climate*, *13*(10), 1748–1761.